\documentclass[a4paper,twocolumn,english,prb,showpacs]{revtex4}
\usepackage[T1]{fontenc}
\usepackage[latin1]{inputenc}
\usepackage{amsmath}
\usepackage{graphicx}
\usepackage{amssymb}

\makeatletter
\usepackage{graphicx}
\usepackage{graphicx}

\usepackage{babel}
\makeatother
\begin{document}

\title{Inter-shell interaction in double walled carbon nanotubes: charge
transfer and orbital mixing}

\author{V. Zólyomi$^{1,2}$, J. Koltai$^{3}$, Á. Rusznyák$^{3}$, J. Kürti$^{3}$,
Á. Gali$^{4}$, F. Simon$^{5}$, H. Kuzmany$^{5}$, Á. Szabados$^{2}$,
and P. R. Surján$^{2}$}

\affiliation{$^{1}$Research Institute for Solid State Physics and Optics of the
Hungarian Academy of Sciences, P. O. B. 49, H-1525, Budapest, Hungary}

\affiliation{$^{2}$Institute for Chemistry, Eötvös University, Pázmány Péter
sétány 1/A, H-1117 Budapest, Hungary}

\affiliation{$^{3}$Department of Biological Physics, Eötvös University, Pázmány
Péter sétány 1/A, H-1117 Budapest, Hungary}

\affiliation{$^{4}$Department of Atomic Physics, Budapest University of Technology
and Economics, Budafoki út 8, H-1111, Budapest, Hungary}

\affiliation{$^{5}$Institut für Materialphysik, Universität Wien, Strudlhofgasse
4, A-1090 Wien, Austria}


\begin{abstract}
Recent nuclear magnetic resonance measurements on isotope engineered
double walled carbon nanotubes (DWCNTs) surprisingly suggest a uniformly
metallic character of all nanotubes, which can only be explained by
the interaction between the layers. Here we study the inter-shell
interaction in DWCNTs by density functional theory and inter-molecular
H\"{u}ckel model. We find charge transfer between the layers using
both methods. We show that not only does the charge transfer appear
already at the fundamental level of the inter-molecular Hückel model,
but also that the spatial distribution of the change in the electron
density is well described already at this level of theory. We find
that the charge transfer between the walls is on the order of 0.001
e/atom and that the inner tube is always negatively charged. We also
observe orbital mixing between the states of the layers. We find that
these two effects combined can in some cases lead to a semiconductor--to--metal
transition of the double walled tube, but not necessarily in all cases.
\end{abstract}
\maketitle

\section{\label{sec:INTRODUCTION}Introduction}

Carbon nanotubes have been intensively studied in the past 15 years
due to their high application potential and their rich physics. Single
walled carbon nanotubes (SWCNTs), in particular, show fundamental
phenomena ranging from e.g. possible superconductivity \cite{BenedictLX_1995_1}
or Luttinger-liquid state \cite{IshiiH_2003_1} to Peierls distortion
\cite{ConnetableD_2005_1}. The electronic properties of SWCNTs are
known to be fully determined by their $(n,m)$ chiral indices (which
essentially define the alignment of the hexagons on the SWCNT surface
with respect to the tube axis) \cite{ThomsenNanobook_2004_1}. Peapod
annealing produced double walled carbon nanotubes (DWCNTs) \cite{BandowS_2001_1}
also possess a number of unique properties such as very long phonon
and optical excitation life-times \cite{PfeifferR_2003_1}. DWCNTs
are interacting systems consisting of two subsystems: an inner and
an outer SWCNT. The subsystems are still well defined by their $(n,m)$
chiral indices, but lose some of their identity due to the interaction,
as suggested by recent experiments. Nuclear magnetic resonance (NMR)
measurements show the extremely surprising result that the DWCNTs
have a highly uniform metallic character \cite{SingerPM_2005_1}.
This observation contradicts theoretical expectations for SWCNTs,
especially in the diameter region of the inner tubes, where curvature
induces a secondary gap in non-armchair tubes that should be metallic
by simple zone folding approximation \cite{ZACHARY_2004_1}. Therefore,
these NMR observations can only be explained by the interaction between
the inner and outer wall. The importance of the interaction is qualitatively
easy to understand compared to the case of bundles, where the interaction
surface of adjacent nanotubes is small, whereas in the case of double
walled carbon nanotubes the interaction surface between the two layers
is 100 \%. Resonant Raman measurements have previously given experimental
evidence for the redshift of the Van Hove transition energies due
to the interaction between the layers in DWCNTs, as well as for a
dependence of the redshift on the inter-shell distance \cite{PfeifferR_2005_1}.

In this work we present the results of our theoretical investigation
of inter-shell interaction and its consequences in DWCNTs. We studied
65 different DWCNTs by inter-molecular Hückel (IMH) model \cite{StafstromIJQC,LaplacePRA}.
We have also studied 6 of these DWCNTs -- 3 commensurate $(n,n)$@$(n',n')$
and 3 commensurate $(n,0)$@$(n',0)$ DWCNTs -- by first principles
density functional theory within the local density approximation (LDA).
We found a semiconductor--to--metal transition in 2 of the 3 $(n,0)$@$(n',0)$
DWCNTs studied by DFT, with only the third one retaining a small band
gap. We have previously reported that our calculations predict a large
density of states at the Fermi-level in the case of metallic non-armchair
DWCNTs, and that starting from two semiconducting SWCNTs the resulting
DWCNTs may transform into a metallic state, but not necessarily in
every case \cite{ZACHARY_2006_2}. In Ref. \cite{ZACHARY_2006_2},
we briefly outlined some of the results of the present paper, namely
that a small charge transfer (CT) from the outer wall to the inner
wall occurs in every DWCNT. This effect has since been confirmed by
photoemission spectroscopy \cite{Hide_2006_1}. Here we present our
results on the charge transfer in full detail. We point out that the
spatial distribution of the change in the electron density according
to the inter-molecular Hückel model is in excellent agreement with
first principles calculations in the case of the most typical inner
diameters, pointing out that the interactions are well described already
at this simple fundamental level of theory. We also find orbital mixing
between the layers, which can explain the measured redshift of the
resonance in the Raman measurements of DWCNTs \cite{PfeifferR_2005_1}.
We conclude that the observed charge transfer and orbital mixing together
can account for a semiconductor--to--metal transition of DWCNTs, but
not necessarily a near-universal metallicity.

\section{Method}

LDA calculations were performed both with a plane wave (VASP \cite{KresseG_1996_2})
and a localized basis set (SIESTA \cite{OrdejonP_SIESTA_1996}) package.
In the VASP calculations the projector augmented-wave method was applied
using a 400 eV plane-wave cutoff energy, while in the SIESTA calculations
double-$\zeta$ plus polarization function basis set was employed.
16 irreducible $k$-points were used; comparison with test calculations
using 31 $k$-points showed this to be sufficient. As these codes
use periodic boundary conditions, only commensurate DWCNTs can be
studied by them in practice. Otherwise, a model of incommensurate
DWCNTs would require huge supercells. An alternate approach to compare
the inter-shell interaction in different DWCNTs is the inter-molecular
Hückel (IMH) model \cite{StafstromIJQC,LaplacePRA}. In this case
the tight binding wave functions originate from the inner and outer
tubes (orbital mixing). Using a Lennard-Jones type expression to account
for inter-cluster interactions, the tight binding model has been applied
to characterize weakly interacting carbon nanotubes \cite{bundlesRBM_Rubio,Tomanekorientationalmelting,MWNT2,TBDonMWNT_Palser}.
The tight binding principle can be generalized to apply to both intra-
and inter-molecular interactions \cite{sven.JMM92,sven.PRB93,sven.PRB99},
leading to the IMH model \cite{StafstromIJQC}, which has been successfully
applied to study DWCNTs \cite{LaplacePRA} and bundles of SWCNTs \cite{Szabados_Biro_PRB06}.
Detailed account of the model is given in Ref. \cite{Szabados_Biro_PRB06}.
The IMH model allows to calculate the charge transfer (CT) for \textit{any}
DWCNT with good efficiency. Test calculations on the commensurate
(7,0)@(16,0) DWCNT show, that the infinite limit is easy to obtain
from calculations on finite DWCNT pieces of gradually increasing length,
and the error is less than 0.2 \%. Furthermore, optimizing the bond
lengths by means of the Longuet-Higgins-Salem model \cite{LHS1959,KeSuSSC81}
prior to the CT calculations, the charge transfer is altered by merely
0.8 \% as compared to the graphene wrapping model, showing that the
CT is not very sensitive to the actual bond lengths at this level
of theory.

\section{\label{sec:Results}Results}

The DWCNTs considered in our calculations were selected based on Raman
measurements \cite{PfeifferR_2005_1}. The experimental diameter distribution
of the outer wall of the DWCNTs was centered at 1.4~nm with a variance
of 0.1~nm while the ideal diameter difference between the inner and
outer wall is 0.72~nm with a variance of 0.05~nm, corresponding
to an inner diameter distribution centered at 0.68~nm. The Raman
measurements clearly show that there is no chirality preference for
the inner-outer tube pairs and a wide range of combinations can be
found in the sample. In accordance with this, we have chosen to examine
the inner tubes which are at the center of the inner diameter distribution,
and to examine each inner tube with various outer tubes, in order
to examine the chirality-dependence of the interactions. In the case
of the IMH calculations, the diameter of each tube was taken from
the usual graphene folding formulas with a uniform bond length of
1.41~\AA, while in the case of the LDA calculations we used optimized
geometries \cite{KurtiJ_2003_1}. All inner tubes with diameters $d_{inner}=0.7\pm0.05$~nm
were examined, and for each inner tube all outer tubes with diameters
$d_{outer}=d_{inner}+0.72\pm0.04$~nm were considered yielding a
total of 60 different DWCNTs. In addition we have also studied five
other DWCNTs which are outside of the aforementioned diameter range,
but could still be present in the sample, in order to compare with
the LDA calculations. The 6 commensurate DWCNTs studied by LDA were:
(4,4)@(9,9), (5,5)@(10,10), (6,6)@(11,11), (7,0)@(16,0), (8,0)@(17,0),
and (9,0)@(18,0).

\begin{figure}[htbp]
\includegraphics[width=8.5cm]{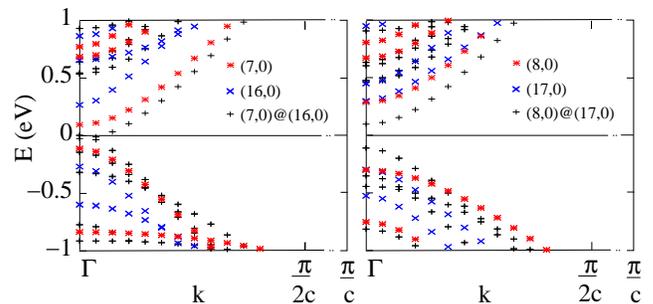}

\caption{\label{cap:bandstructs}LDA band structures of the (7,0)@(16,0) and
(8,0)@(17,0) DWCNTs, in comparison with the band structures of their
subsystems in isolated single geometry (the Fermi levels are all shifted
to 0 eV). }
\end{figure}

We calculated the band structure of the 6 commensurate DWCNTs by both
LDA packages, and found fairly good agreement between localized basis
set calculations and well-converged plane wave results. The three
armchair DWCNTs are all metallic, exactly as expected. Of the three
zigzag DWCNTs, (8,0)@(17,0) remains a semiconductor, while the other
two are metallic. Note, that all zigzag SWCNTs considered were originally
semiconducting: the LDA gaps of (7,0) and (16,0) were 0.21 eV and
0.54 eV, respectively, while those of (9,0) and (18,0) were 0.096
eV and 0.013 eV, respectively \cite{ZACHARY_2004_1}. The LDA band
gap of (8,0)@(17,0) is about 0.2~eV which is much smaller than that
of the individual SWCNTs of about 0.6~eV \cite{ZACHARY_2004_1}.
The band structures of the (7,0)@(16,0) and (8,0)@(17,0) DWCNTs are
plotted in Figure \ref{cap:bandstructs}. 

\begin{figure}[htbp]
\includegraphics[width=8.5cm]{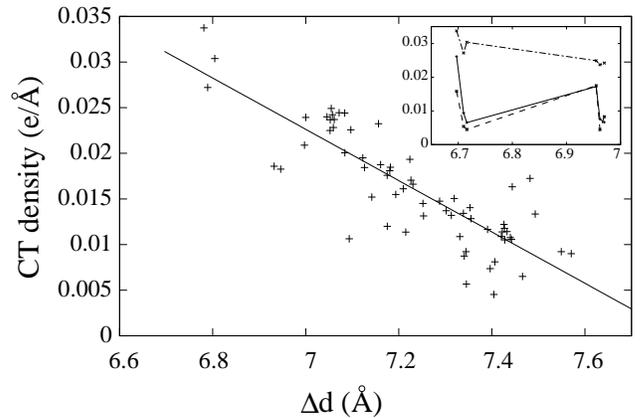}

\caption{\label{cap:CTpic}Charge transfer density along the tube axis versus
the diameter difference ($\Delta d$) between the inner and outer
tube according to the IMH model. The straight line is a linear $a\cdot\Delta d+b$
regression (see text). The inset shows a comparison between IMH and
LDA results. The CT values obtained from the VASP calculations by
analyzing the Bader charges (dashed) agree well with those obtained
from the SIESTA calculations by Mulliken population analysis (solid),
but both of them give a smaller value than the IMH method (dashed-dotted).
The 6 points of each calculation correspond to (4,4)@(9,9), (5,5)@(10,10),
(6,6)@(11,11), (7,0)@(16,0), (8,0)@(17,0), and (9,0)@(18,0), from
left to right.}
\end{figure}

An earlier study of a linear carbon chain in SWCNT \cite{ZACHARY_2005_1}
indicates that metallicity may occur due to charge transfer between
the layers of the DWCNT. To investigate the reasons for metallicity
in more detail, we have calculated the CT in the 65 DWCNTs in the
IMH model. This is straightforwardly determined by summing up the
contributions of the LCAO-coefficients of every occupied molecular
orbital separately for the atoms in the inner and outer wall, and
then comparing with the number of electrons that should be present
on the given wall if there was no CT. Figure \ref{cap:CTpic} shows
our results for the charge transfer density along the tube axis as
a function of the difference between the inner and outer diameter
($\Delta d=d_{outer}-d_{inner}$). In all cases, we found that the
inner tubes are negatively charged. This result is in perfect agreement
with recent observations of photoemission spectra of DWCNTs, which
also predict negatively charged inner tubes \cite{Hide_2006_1}. Our
calculated values for the CT density were between 0.005 e/\AA~ and
0.035 e/\AA. This corresponds to a range of about 0.0005 to 0.0045
e/atom for the inner wall, and 0.0002 to 0.0024~e/atom for the outer
wall; note, that this CT is much smaller than what is typical in e.g.\ 
alkali-intercalation experiments. While the CT values show a decent
amount of scattering, there is also a strong and clear overall decrease
of the CT as $\Delta d$ increases, which is expected, as the overlap
between the orbitals of the two separate layers decreases as the distance
between them increases. The large variance of the points is related
to the difficulty of accurately predicting a CT of this order of magnitude.
However, it is safe to conclude, that the CT density in units of e/\AA~
can be estimated by the linear formula $-0.028\cdot\Delta d+0.219$
with a considerable variance depending on tube chirality.

Advancing beyond the IMH model, we calculated the charge transfer
for the six commensurate DWCNTs by LDA. In the case of the plane-wave
calculation, were were able to calculate the Bader-charges with Voronoi
partitions (which defines the borders of the atomic volumes by planes
half way between atoms, similar to the construction of Wigner-Seitz
cells) using an external utility \cite{HenkelmanG2006_1}. In the
case of the localized basis set calculation, Mulliken population analysis
could be performed. We found, that the direction and the order of
magnitude of the CT is the same in these cases as what was found with
IMH. The LDA CT is however somewhat smaller. This is illustrated in
the inset of Figure \ref{cap:CTpic}. Note, that the Bader charges
agree very well with the Mulliken charges. This is not necessary to
occur, as the Mulliken population analysis usually only performs well
in minimal basis set, however, if all atoms are of the same species
(like in our case) then a larger basis set can also give reliable
Mulliken charges, and apparently such is the case in our calculations. 

Based on the good agreement between the three different CT calculations,
we conclude that in DWCNTs electrons are transferred from the outer
to the inner wall on the order of 0.005 to 0.035~e/\AA, depending
on tube chiralities.

\section{\label{sec:Discussion}Discussion}

We found for the 3 commensurate $(n,0)$@$(n',0)$ DWCNTs that, due
to the small magnitude of the charge transfer, the Fermi level is
close to the Van Hove singularities that used to form the band gap
of the single walled subsystems. This results in a large density of
states at the Fermi level \cite{ZACHARY_2006_2}, very similar to
the case of chain@SWCNT systems \cite{ZACHARY_2005_1}. Based on our
IMH results, which say that the chirality-dependence of the magnitude
of the CT is more-less uniform, we can safely conclude that this large
density of states at the Fermi level is expected in all of those metallic
DWCNTs, where at least one of the two subsystems is a non-armchair
tube, as all but armchair tubes have Van Hove singularities near the
Fermi level at diameters above 0.5~nm \cite{ZACHARY_2004_1}. Thus
we expect that the majority of the metallic DWCNTs have a large density
of states at the Fermi level. This behavior is very similar to doped
multi-walled tubes, which have previously been suggested as possible
future superconductors \cite{BenedictLX_1995_1}. 

The case of the non-metallic (8,0)@(17,0) DWCNT deserves attention.
The two subsystems are semiconducting having almost the same band
gap ($\approx0.6$ eV) at LDA level \cite{ZACHARY_2004_1}. The DWCNT
they form remains semiconducting, but the bands near the Fermi level
are distorted as compared to the rigid band prediction, such that
the band gap drops to $\approx0.2$~eV. This result underlines the
importance of orbital mixing, and points out that the interaction
between the inner and outer tubes is not limited merely to charge
transfer, but the mixing of inner and outer tube orbitals is also
an important part of the interaction. Furthermore, this result also
shows that the orbital mixing caused by the inter-shell interaction
provides the explanation to the experimentally observed redshift of
the Van Hove transition energies \cite{PfeifferR_2005_1}. The redshift
is immediately understood by the contraction of the bands such as
in the case of (8,0)@(17,0) in Figure \ref{cap:bandstructs}. In the
experiments, all Van Hove transition energies show a redshift, with
the lower energy transitions of a DWCNT suffering a greater shift
than its higher energy transitions; our calculations show exactly
the same qualitative trend.

Thus, from the point of view of the electronic states, a DWCNT should
-- strictly speaking -- always be considered as one single unified
system. Approximating a DWCNT by separating it to an inner and an
outer subsystem is definitely possible, but it should always be done
with caution. For example, if orbital mixing were small, one could
estimate whether a given SWCNT is likely to become metallic as one
layer of a DWCNT, by calculating the critical charge transfer (CT$_{crit}$)
-- the CT where the isolated tube becomes metallic upon doping --
for the charged, isolated SWCNT. However, this method is not reliable
for DWCNTs, because it completely neglects orbital mixing, which is
obviously an important factor, as detailed above. In fact, we have
calculated CT$_{crit}$ for the two subsystems of the (8,0)@(17,0)
DWCNT with this method, and in both cases we obtained a value which
is about a factor of 2 \textit{smaller} than the CT from the DWCNT
calculation. This contradictory result clearly shows that the question
of whether a given DWCNT is metallic cannot be answered by means of
calculating CT$_{crit}$ on charged SWCNTs.

Another reason why it is not trivial to separate the two subsystems
is the delocalized profile of the spatial distribution of the change
of the electron density induced by the inter-layer interaction, which
we will refer to as the charge redistribution profile. It has been
pointed out in previous works in the local density approximation,
that upon examining the contour plots of the change of the electron
density in DWCNTs, it could be found that the electrons deplete from
the walls and accumulate in the space between the layers \cite{MiyamotoY_2002_1,OkadaS_2003_1,ShanB_2006_1}.
We have also performed these calculations on the charge redistribution
profile for the DWCNTs we examined, and found good agreement with
these previous works (see below). These results suggest that it is
very difficult to divide the charges of the total charge density between
the layers. Experiments however show that the layers behave more-less
individually, as e.g. Van Hove transitions of the inner and outer
tubes can be clearly identified in Raman measurements \cite{PfeifferR_2005_1}.
Furthermore, recent measurements clearly identify an inter-layer charge
transfer in DWCNTs \cite{Hide_2006_1}, as mentioned earlier. We have
used 3 different methods to calculate the charge transfer between
the layers, and all 3 methods showed good agreement with the experimentally
observed direction of the charge transfer. This shows that the charge
transfer analyses we conducted are able to perform a plausible separation
of the charge density between the inner and outer nanotube.

\begin{figure}[htbp]
\includegraphics[width=8.5cm]{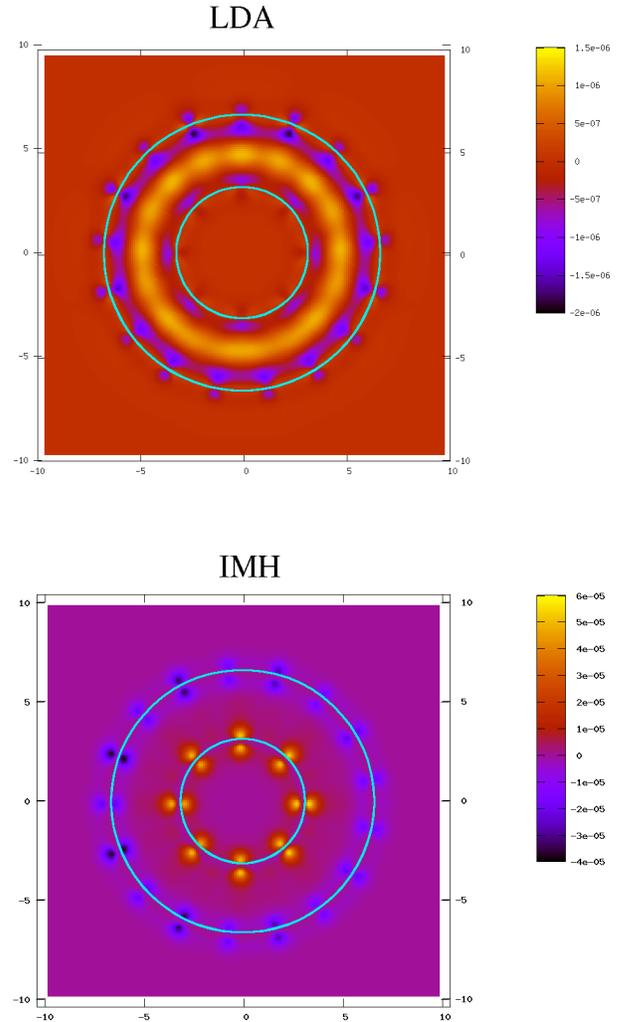}

\caption{\label{cap:palette_1}Charge redistribution profile (change of the
electron density caused by inter-layer interaction) in the case of
the (8,0)@(17,0) DWCNT, in one of the planes perpendicular to the
tube axis containing the atoms (units are electrons per \AA$^{3}$).
The LDA results also agree well with previous calculations \cite{ShanB_2006_1},
while the IMH results show a more localized redistribution profile
owing to the neglection of $s-p$ mixing (see text).}
\end{figure}

\begin{figure}[htbp]
\includegraphics[width=8.5cm]{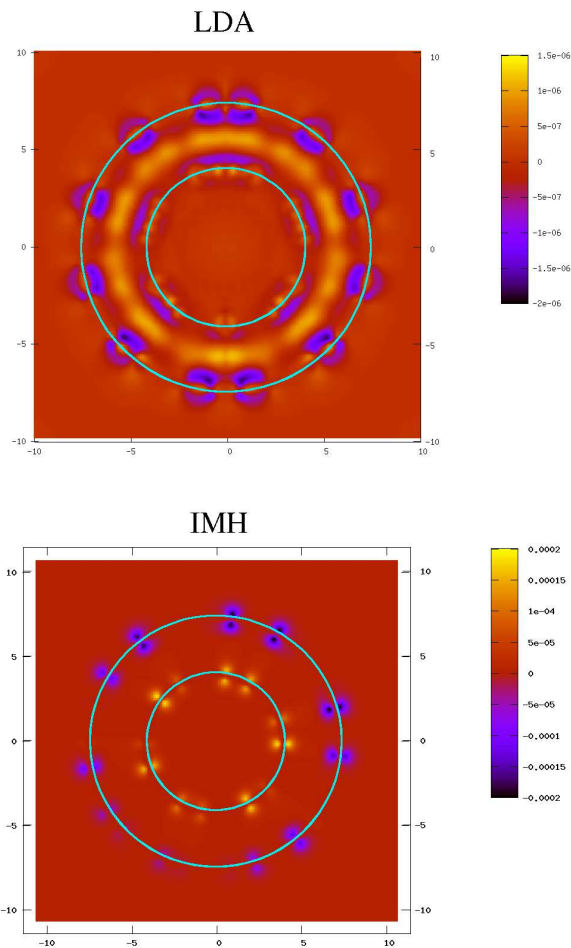}

\caption{\label{cap:palette_2}Charge redistribution profile (change of the
electron density caused by inter-layer interaction) in the case of
the (6,6)@(11,11) DWCNT, in one of the planes perpendicular to the
tube axis containing the atoms (units are electrons per \AA$^{3}$). }
\end{figure}

Finally, in order to make a further comparison between the LDA and
IMH results, we have calculated the charge redistribution profile
using IMH as well. As mentioned above, the LDA results for the (8,0)@(17,0)
are in good agreement with previous calculations \cite{ShanB_2006_1}. 

Our results are plotted in Figures \ref{cap:palette_1} and \ref{cap:palette_2},
showing the comparison between IMH and LDA in the case of the (8,0)@(17,0)
and (6,6)@(11,11) DWCNTs. The IMH model shows a different redistribution
profile than the LDA calculation, showing a picture that the electrons
deplete from the outer tube and accumulate on the inner wall. This
result shows, that while the IMH model fails to reproduce the correct
spatial charge redistribution profile for this DWCNT, it qualitatively
agrees with our result on the direction of the CT. Thus we conclude
that proper inclusion of $s-p$ mixing is necessary in order to arrive
at the correct spatial distribution of the charge density at small
diameters.

\section{\label{sec:Conclusions}Conclusion}

In conclusion, we have performed calculations on a large number of
double walled carbon nanotubes, using density functional theory and
the inter-molecular Hückel model. We have found that electrons are
transferred from the outer tube to the inner tube in all cases, with
the magnitude of the average charge transfer density along the tube
axis ranging from 0.005 e/\AA~ to 0.035 e/\AA, depending on tube
chiralities. We have found that inter-layer orbital mixing is a very
important part of the interaction between the layers, and that the
interactions can turn a DWCNT consisting of semiconducting subsystems
into a metal, but not necessarily in every case. We predict that the
majority of metallic DWCNTs have a high density of states at the Fermi
level. We have also found that the charge redistribution profile is
qualitatively different in the IMH and the LDA calculations due to
the neglection of $s-p$ mixing in the former method. 

$ $

\begin{acknowledgments}
\label{ack:Support-from-OTKA}We thank O. Dubay and R. Pfeiffer for
valuable discussions. Support from OTKA (Grants No. T038014, K60576,
T43685, T49718, D45983, TS049881, F61733, and NK60984) in Hungary,
the Austrian science foundation (project 17345), and the MERG-CT-2005-022103
EU project is gratefully acknowledged. V. Z. also acknowledges the
EU project NANOTEMP BIN2-2001-00580, and F. S. also acknowledges the
Zoltán Magyary program.

\bibliographystyle{apsrev}
\bibliography{/home/zachary/PHYS_ALL/bibtex_file/article_list}

\end{acknowledgments}

\end{document}